\documentclass[prd,eqsecnum,preprint,showpacs]{revtex4}
\usepackage{amsmath}
\usepackage[dvips]{graphicx}
\input amssym

\DeclareMathOperator{\atgh}{arctanh}
\DeclareMathOperator{\diag}{diag}
\DeclareMathOperator{\Li}{Li}

\begin{document}

\def\prg#1{\medskip\noindent{\bf #1}}  \def\ra{\rightarrow}
\def\lra{\leftrightarrow}              \def\Ra{\Rightarrow}
\def\nin{\noindent}                    \def\pd{\partial}
\def\dis{\displaystyle}                \def\inn{\,\rfloor\,}
\def\grl{{GR$_\Lambda$}}               \def\vsm{\vspace{-9pt}}
\def\Lra{{\Leftrightarrow}}
\def\cs{{\scriptstyle\rm CS}}          \def\ads3{{\rm AdS$_3$}}
\def\Leff{\hbox{$\mit\L_{\hspace{.6pt}\rm eff}\,$}}
\def\bull{\raise.25ex\hbox{\vrule height.8ex width.8ex}}
\def\ric{{(Ric)}}                      \def\tric{{(\widetilde{Ric})}}
\def\tmgl{\hbox{TMG$_\Lambda$}}
\def\Lie{{\cal L}\hspace{-.7em}\raise.25ex\hbox{--}\hspace{.2em}}
\def\sS{\hspace{2pt}S\hspace{-0.83em}\diagup}   \def\hd{{^\star}}
\def\dis{\displaystyle}

\def\hook{\hbox{\vrule height0pt width4pt depth0.3pt
\vrule height7pt width0.3pt depth0.3pt
\vrule height0pt width2pt depth0pt}\hspace{0.8pt}}
\def\semidirect{\;{\rlap{$\supset$}\times}\;}
\def\first{\rm (1ST)}
\def\second{\hspace{-1cm}\rm (2ND)}
\def\bm#1{\hbox{{\boldmath $#1$}}}
\def\nb#1{\marginpar{{\large\bf #1}}}

\def\G{\Gamma}        \def\S{\Sigma}        \def\L{{\mit\Lambda}}
\def\D{\Delta}        \def\Th{\Theta}
\def\a{\alpha}        \def\b{\beta}         \def\g{\gamma}
\def\d{\delta}        \def\m{\mu}           \def\n{\nu}
\def\th{\theta}       \def\k{\kappa}        \def\l{\lambda}
\def\vphi{\varphi}    \def\ve{\varepsilon}  \def\p{\pi}
\def\r{\rho}          \def\Om{\Omega}       \def\om{\omega}
\def\s{\sigma}        \def\t{\tau}          \def\eps{\epsilon}
\def\nab{\nabla}      \def\btz{{\rm BTZ}}   \def\heps{\hat\eps}

\def\tG{{\tilde G}}   \def\cF{{\cal F}}
\def\cL{{\cal L}}     \def\cM{{\cal M }}   \def\cE{{\cal E}}
\def\cH{{\cal H}}     \def\hcH{\hat{\cH}}
\def\cK{{\cal K}}     \def\hcK{\hat{\cK}}  \def\cT{{\cal T}}
\def\cO{{\cal O}}     \def\hcO{\hat{\cal O}} \def\cV{{\cal V}}
\def\tom{{\tilde\omega}}  \def\cE{{\cal E}}
\def\cR{{\cal R}}    \def\hR{{\hat R}{}}   \def\hL{{\hat\L}}
\def\tb{{\tilde b}}  \def\tA{{\tilde A}}   \def\tv{{\tilde v}}
\def\tT{{\tilde T}}  \def\tR{{\tilde R}}   \def\tcL{{\tilde\cL}}

\def\nn{\nonumber}                    \def\vsm{\vspace{-8pt}}
\def\be{\begin{equation}}             \def\ee{\end{equation}}
\def\ba#1{\begin{array}{#1}}          \def\ea{\end{array}}
\def\bea{\begin{eqnarray} }           \def\eea{\end{eqnarray} }
\def\beann{\begin{eqnarray*} }        \def\eeann{\end{eqnarray*} }
\def\beal{\begin{eqalign}}            \def\eeal{\end{eqalign}}
\def\lab#1{\label{eq:#1}}             \def\eq#1{(\ref{eq:#1})}
\def\bsubeq{\begin{subequations}}     \def\esubeq{\end{subequations}}
\def\bitem{\begin{itemize}}           \def\eitem{\end{itemize}}

\title{Gravitational waves with torsion in 3D}

\author{M. Blagojevi\'c\,}\email{mb@ipb.ac.rs}
\affiliation{University of Belgrade, Institute of Physics,
P. O. Box 57, 11001 Belgrade, Serbia}
\author{B. Cvetkovi\'c\,}\email{cbranislav@ipb.ac.rs}
\affiliation{University of Belgrade, Institute of Physics,
P. O. Box 57, 11001 Belgrade, Serbia}
\date{\today}

\begin{abstract}
We study gravitational waves with torsion as exact vacuum solutions of
three-dimensional gravity with propagating torsion. The new solutions are
a natural generalization of the plane-fronted gravitational waves in
general relativity with a cosmological constant, in the presence of
matter.
\end{abstract}
\maketitle
\section{Introduction}
\setcounter{equation}{0}

Investigations of three-dimensional (3D) gravity have had an important
influence on our understanding of both classical and quantum aspects of
the realistic gravitational dynamics. In this context, the traditional
approach based on \emph{general relativity} has led to a number of
outstanding results \cite{x1}. However, in the early 1990s, Mielke and
Baekler \cite{x2} initiated a new approach to 3D gravity, relying on a
modern field-theoretic formulation of gravity, the \emph{Poincar\'e gauge
theory} (PGT), proposed in the early 1960s by Kibble and Sciama
\cite{x3,x4,x5,x6}. Compared to general relativity, dynamical structure of
PGT is extended by using both the curvature and the torsion to describe
the associated Riemann--Cartan (RC) geometry of spacetime.

The Mielke--Baekler model, like Einstein's general relativity, is a
topological theory without propagating degrees of freedom. In PGT, such an
unrealistic feature of the gravitational dynamics can be naturally
improved by going over to a Lagrangian that is at most quadratic in
torsion and curvature (quadratic PGT). Recent investigations reveal
elements that indicate a rich dynamical structure of the \emph{quadratic
PGT}\, \cite{x7,x8,x9,x10}: the theory possesses a number of
\emph{propagating torsion modes} (tordions) and black hole solution, its
(A)dS sector is characterized by well-defined conserved charges and
central charges, the existence of torsion is compatible with the AdS/CFT
correspondence, and the canonical structure shows a close resemblance with
the four-dimensional theory.

In the present paper, we continue studying dynamical aspects of the
quadratic PGT in 3D by looking for \emph{exact wave solutions with
torsion}. The weak-field approximation of Einstein's theory around the
Minkowski background leads to a simple picture of the wave nature of
gravity, which is recognized to have a striking analogy to the
electromagnetic phenomena \cite{x11,x12}. By giving a covariant
formulation of this analogy, one can generalize the linearized
gravitational wave to the concept of an exact wave solution of general
relativity \cite{x13,x14,x15}. Here, in the context of the quadratic PGT,
such generalizations are used to find a class of exact wave solutions with
torsion.

A gravitational wave with torsion in 3D was first found by Obukhov
\cite{x16}, in the framework of the Mielke--Baekler model \cite{x2}. Since
the model is defined by a topological action, it was necessary to
introduce matter, chosen in the form of Maxwell field, to have a
nontrivial wave solution. On the other hand, our wave solution, being an
exact vacuum solution of the quadratic PGT, offers a new insight into the
wave structure of genuine gravitational degrees of freedom, the
propagating torsion modes.

The paper is organized as follows. In section 2, we give an overview of
the plane-fronted gravitational waves in general relativity without/with
gravitational constant, denoted shortly as GR/\grl, as a basis for further
extension to torsion waves in the quadratic PGT. In section 3, we start
with the \grl\  form of the metric and introduce a convenient ansatz for
the RC connection, or equivalently, for the torsion. The only irreducible
component of torsion is taken to be its tensorial piece, parametrized by a
single function $K$. Then, we find the PGT field equations that impose
dynamical restrictions on $K$. A characteristic parameter appearing in
these equations is the mass parameter $\m^2$, associated to the torsion
spin-2 mode. In sections 4 and 5, we find a class of exact torsion waves
and classify them according to the values of two parameters, $\m^2$ and
$\l$, the latter one being related to the value of cosmological constant.
In section 6, we discuss criteria that are used to recognize the wave
nature of exact solutions and conclude with some specific remarks.
Finally, two appendices contain a useful technical information.

Our conventions are the same as in Ref. \cite{x8}: the Latin indices $(i,
j,k, ...)$ refer to the local Lorentz frame, the Greek indices $(\m, \n,
\r, ...)$ refer to the coordinate frame, and both run over 0,1,2; the
metric components in the local Lorentz frame are $\eta_{ij}= (+, -, -)$;
totally antisymmetric tensor $\ve^{ijk}$ is normalized to $\ve^{012} = 1$;
$b^i$ is the orthonormal triad (coframe 1-form), $h_i$ is the dual basis
(frame), the Hodge dual of a form $\a$ is $\hd \a$, and the exterior
product of forms is implicit.

\section{Plane-fronted waves in general relativity}
\setcounter{equation}{0}

In this section, we give  a short account of the plane-fronted
gravitational waves as exact solutions of Einstein's general relativity.

\subsection{pp-waves in GR}

A specific class of plane-fronted waves, characterized by having parallel
rays (pp--waves for short), can be described, in suitable local
coordinates, by the metric \cite{x13,x14,x15}
\be
ds^2=H(u,y)du^2+2du dv-dy^2\, ,                                 \lab{2.1}
\ee
where $u$ is interpreted as the phase of the wave and $\pd_v$ is the
covariantly constant null vector field. This metric is a natural
generalization of the \emph{linearized} gravitational plane waves
propagating on the background Minkowski spacetime \cite{x11,x12}. General
criteria for identifying the wave nature of \emph{exact} solutions will be
discussed in section \ref{Dis}.

The explicit form of $H(u,y)$ in \eq{2.1} can be determined by the GR
field equations. Since the only nonvanishing component of the Ricci tensor
is $\ric_{uu}=H''/2$ (prime means differentiation with respect to $y$) and
the scalar curvature identically vanishes, $R=0$, the \emph{vacuum} field
equations of GR imply
\be
H''=0\quad\Ra\quad H=h_1(u)+h_2(u)y\, ,                       \lab{2.2}
\ee
where $h_1,h_2$ are the integration ``constants". This solution is in fact
trivial since for $H''=0$ the Ricci tensor vanishes and, in 3D, the full
curvature tensor also vanishes. Hence, \eq{2.2} defines a Minkowski
spacetime in nonstandard coordinates.

Thus, in GR, nontrivial pp-waves can exist only in the presence of
\emph{matter}; see, for instance \cite{x17,x18,x19}. Note, however, that
true \emph{vacuum} waves can exist also in \emph{new dynamical settings},
such as Topologically massive gravity or New massive gravity
\cite{x19,x20,x21}. The vacuum waves are an idealization of wave solutions
in the region far from matter sources.

\subsection{Plane-fronted waves in \grl}

Now, we turn to a generalized dynamical framework of \grl\ by allowing a
nonvanishing cosmological constant. The pp-wave \eq{2.1} is not a vacuum
solution of \grl. Indeed, the fact that $R=0$ for the metric \eq{2.1}
implies $\L=0$. A plane-fronted wave that is compatible with $\L\ne 0$ can
be conveniently represented by the metric
\bsubeq\lab{2.3}
\be
ds^2=2\left(\frac{q}{p}\right)^2du(Sdu+dv)-\frac{dy^2}{p^2}\, ,
\ee
see Ozsv\'ath \cite{x22} and Obukhov \cite{x23}, where the functions $p,q$
and $S$ are chosen as \cite{x16}
\be
p=1+\frac{\l}{4}y^2\, ,\qquad q=1-\frac{\l}{4}y^2\,,\qquad
S=-\frac{\l}{2}v^2+\frac{\sqrt{p}}{2q}H(u,y)\,.
\ee
\esubeq
Clearly, the limit $\l=0$ returns us back to the pp-wave \eq{2.1}.
Introducing the ortonormal triad field as
\bea
&& b^0:=\frac{1}{\sqrt{2}}\left[\left(1+\frac{q^2}{p^2}S\right)du
                                +\frac{q^2}{p^2}dv\right]\, ,   \nn\\
&& b^1:=\frac{1}{\sqrt{2}}\left[\left(1-\frac{q^2}{p^2}S\right)du
                                -\frac{q^2}{p^2}dv\right]\, ,   \nn\\
&& b^2:=\frac{1}{p}dy\, ,                                       \lab{2.4}
\eea
the metric can be written as $ds^2=\eta_{ij}b^i\otimes b^j$, with
$\eta_{ij}=\diag (+1,-1,-1)$. In the literature, one often uses the
light-cone components of the triad:
$$
b^+:=du\,,\qquad b^-:=\frac{q^2}{p^2}(Sdu+dv)\,.
$$

In order to verify that that the triad \eq{2.4} satisfies the \grl\ field
equations,
\be
a_0\left(\ric^i-\frac{1}{2}Rb^i\right)-\L b^i=0\, ,\qquad
a_0:=\frac{1}{16\pi G}\, ,                                      \lab{2.5}
\ee
we first calculate the Christofell connection; it has the form
\bea
&&\G^{01}=\frac{\l y}{q}b^2-\frac{\l v}{\sqrt{2}}(b^0+b^1)\, ,\nn\\
&&\G^{02}=\frac{\l y}{q}b^0-\frac{1}{2}(b^0+b^1)(q^2S'/p)\, , \nn\\
&&\G^{12}=\frac{\l y}{q}b^1+\frac{1}{2}(b^0+b^1)(q^2S'/p)\, , \nn
\eea
or, more compactly:
\be
\G^{ij}=\bar\G^{ij}
          +\frac{1}{2}\ve^{ij}{_m}k^m k_n b^n(q^2S'/p)\,.       \lab{2.6}
\ee
Here, the first term, $\bar\G^{ij}:=\G^{ij}(S'=0)$, is the piece that
describes the ``background" (A)dS geometry of spacetime, whereas the
second term is the radiation piece, characterized by the null vector
$k^i=(1,-1,0)$, $k^2=0$, which is not covariantly constant for $\l\ne 0$.

Next, we calculate the curvature $R^{ij}=d\G^{ij}+\G^i{_m}\G^{mj}$ ,
\bsubeq\lab{2.7}
\be
R^{ij}=-\l b^i b^j+\ve^{ijm}k_m k^n\hd b_n p(q^2 S'/p)'\, ,     \lab{2.7a}
\ee
where $\hd b_n=(1/2)\ve_{nrs}b^rb^s$. Note that the radiation piece of
$R^{ij}$ is clearly separated from the (A)dS piece. Finally, the form of
the Ricci curvature $\ric^i=-h_j\hook R^{ij}$ and the scalar curvature
$R=h_i\hook\ric^i$,
\bea
&&\ric^i=-2\l b^i+\frac{1}{2}k^ik_mb^m p(q^2S'/p)'\,,           \nn\\
&&R=-6\l\, ,                                                    \lab{2.7b}
\eea
\esubeq
implies that the content of the field equations \eq{2.5} is given by
\bea
&&a_0\l=\L\, ,\qquad p\left(\frac{q^2}{p}S'\right)'=0\, ,       \lab{2.8}\\
&&\Ra\quad\frac{\sqrt{p}}{2q}H=\b_1(u)+\b_2(u)\frac{y}{q}\, .   \nn
\eea
The function $H$ defines the vacuum solution for the metric \eq{2.3}.
Since the on-shell value of the curvature is $R^{ij}=-\l b^ib^j$, the
geometry of the solution \eq{2.8} is fixed: for $\l=0$, $>0$, or $<0$, it
has the Minkowskian, AdS, or dS form, respectively.

Thus, again, in order for the plane-fronted wave \eq{2.3} to be a
nontrivial exact solution, one has to introduce matter. However, by going
over to PGT, we expect the new gravitational dynamics to allow for the
existence of true wave solutions even in vacuum.

\section{Dynamics of torsion waves}
\setcounter{equation}{0}

In this section, we briefly recapitulate basic aspects of PGT, introduce a
geometric extension of the Riemannian plane-fronted waves \eq{2.3} to
torsion waves, and discuss their dynamics.

\subsection{Basic aspects of PGT}

The PGT is a gauge theory of gravity based on gauging the Poincar\'e
group, with an underlying Riemann--Cartan (RC) geometry of spacetime
\cite{x4,x5,x6}. Basic gravitational variables are the triad field $b^i$ and
the Lorentz connection $A^{ij}=-A^{ji}$ (1-forms), and the corresponding
field strengths are the torsion $T^i=db^i+A^i{_k}b^k$ and the curvature
$R^{ij}=d A^{ij}+A^i{_k}A^{kj}$ (2-forms). General dynamics of PGT is
defined by the gravitational Lagrangian $L_G=L_G(b^i,T^i,R^{ij})$
(3-form). Varying $L_G$ with respect to $b^i$ and $A^{ij}$ yields the
respective gravitational field equations in vacuum \cite{x8},
\bea
\first\quad &&\nab H_i+E_i=0\,,                                 \nn\\
\second\quad &&\nab H_{ij}+E_{ij}=0\, ,                         \lab{3.1}
\eea
where
$$
H_i:=\frac{\pd L_G}{\pd T^i}\,,\qquad H_{ij}:=\frac{\pd L_G}{\pd R^{ij}}\,,
$$
are the covariant field momenta, and
$$
E_i:=\frac{\pd L_G}{\pd b^i}\,,\qquad E_{ij}:=\frac{\pd L_G}{\pd A^{ij}}\,,
$$
are the gravitational energy-momentum and spin currents. We require $L_G$
to be parity invariant and at most quadratic in the field strengths. In
that case, $H_i$ and $H_{ij}$ can be expressed linearly in terms of the
irreducible pieces of the field strengths (Appendix A),
\bsubeq\lab{3.2}
\bea
&&H_i=2\hd\left(
     a_1{}^{(1)}T_i+a_2{}^{(2)}T_i+a_3{}^{(3)}T_i\right)\, ,    \nn\\
&&H_{ij}=-2a_0\ve_{ijk}b^k+H'_{ij}\, ,                          \nn\\
&&H'_{ij}:=2\hd\left(b_4{}^{(4)}R_{ij}
           +b_5{}^{(5)}R_{ij}+b_6{}^{(6)}R_{ij}\right)\, ,
\eea
where $a_0,a_n$ and $b_n$ are coupling constants; moreover, the
gravitational Lagrangian takes the form
\be
L_G=\frac{1}{2}T^i H_i+R^{ij}(-a_0\ve_{ijk}b^k)
    +\frac{1}{4}R^{ij}H'_{ij}-\frac{1}{3}\L_0\ve_{ijk}b^ib^jb^k\, ,
\ee
and the gravitational energy-momentum and spin currents turn out to be
\bea
&&E_i=h_i\hook L_G-(h_i\hook T^m)H_m+\frac{1}{4}(h_i\hook R^{mn})H_{mn}\,,\nn\\
&&E_{ij}=-(b_iH_j-b_jH_i)\, .
\eea
\esubeq

\subsection{Geometry of the ansatz}

In our search for the generalized plane-fronted waves, we assume that the
form of the triad field \eq{2.4} remains unchanged, whereas the connection
is determined by the following rule:
\bitem
\item[\bull\,] Starting with the Riemannian connection \eq{2.6}, (i) we
    leave its first, (A)dS piece $\bar\G^{ij}$ unchanged, (ii) but modify
    the second, radiation piece in a way that \emph{preserves} the wave
    nature of the solution.
\eitem
The instruction (ii) is realized by adopting the following ansatz for the
RC connection:
\bsubeq\lab{3.3}
\bea
&&A^{ij}=\bar\G^{ij}+\frac{1}{2}\ve^{ij}{_m}k^m k_n b^n G\, ,   \lab{3.3a}\\
&&G:=\frac{q^2}{p}(S'+K)\, .                                    \lab{3.3b}
\eea
\esubeq
Here, the new term $K=K(u,y)$ describes the effect of torsion, as follows
from
\be
T^i:=\nab b^i=\frac{q^2}{2p}Kk^ik_m\hd b^m\, .                  \lab{3.4}
\ee
The only nonvanishing irreducible piece of $T^i$ is its tensorial piece
(Appendix A):
$$
{}^{(1)}T^i=T^i\, .
$$

Having chosen the form of the connection, one can now calculate the RC
curvatures; they are obtained from \eq{2.7} by the replacement $S'\to
S'+K$:
\bea
&&R^{ij}=-\l b^ib^j+\ve^{ijm}k_m k^n\hd b_n\, pG' \, ,          \nn\\
&&\ric^i=-2\l b^i+\frac{1}{2}k^ik_mb^m pG'\,,                   \nn\\
&&R=-6\l\, .                                                    \lab{3.5}
\eea
The nonvanishing irreducible components of the curvature $R^{ij}$ are
(Appendix A)
$$
{}^{(4)}R^{ij}=\frac{1}{2}\ve^{ijm}k_m k^n\hd b_n\, pG'\, ,\qquad
{}^{(6)}R^{ij}=-\l b^ib^j\, ,
$$
and the quadratic curvature invariant has the form
$R^{ij}\,{}^*R_{ij}=6\l^2\hd 1$.

The geometric configuration defined by the triad field \eq{2.4} and the
connection \eq{3.3} represents a generalized gravitational plane-fronted
wave of \grl, or the \emph{torsion wave} for short. More details on its
wave nature  will be given in section \ref{Dis}.

\subsection{Field equations}

Having found the expressions for the torsion and the curvature, one can
now calculate the covariant momenta $H_i,H_{ij}$, and the energy-momentum
and spin currents $E_i,E_{ij}$, and obtain the explicit form of the PGT
field equations \eq{3.1}. The result takes the following form \cite{x24}:
\bea
(1ST)\quad &&(a_0+b_4\l+b_6\l)pG'-a_1q(qK)'=0\, ,               \nn\\
            &&2\L-2a_0\l+b_6\l^2=0\, ,                          \nn\\
(2ND)\quad &&b_4(2G''p^3q+G'\l y p^3+2G'\l y p^2 q)
             +2(a_1-a_0-b_6\l)Kq^3=0\, .                        \lab{3.6}
\eea
The second equation in $(1ST)$ defines a relation between the parameter
$\l$ of the solution and the coupling constants. For $b_6=0$, it
takes a particularly simple form: $a_0\l=\L$. By noting that $(2ND)$ can
be rewritten as
$$
2b_4p\bigl[pq(pG')'+(pG')\l y\bigr]+2(a_1-a_0-b_6\l)Kq^3=0\, ,
$$
one finds that the field equations \eq{3.6} can be transformed to a more
compact form:
\bea
&&(1ST)\qquad pG'=C_0q\cK'\,,\qquad C_0=\frac{a_1}{a_0+(b_4+b_6)\l}\,,\nn\\
&&(2ND)\qquad  p(p\,\cK')'+\m^2\cK=0\,,
               \qquad \m^2=\frac{a_1-a_0-b_6\l}{b_4 C_0}\, ,    \lab{3.7}
\eea
with $\cK:=qK$.

In PGT, the spectrum of excitations around the Minkowski spacetime
consists of 6 independent torsion modes: one scalar, one pseudoscalar, two
spin-$1$ and two spin-$2$ states \cite{x7,x8}. Two spin-$2$ states form a
parity invariant multiplet associated to the tensorial piece of the
torsion, with equal masses: $m^2=a_0(a_1-a_0)/(a_1b_4)$. Since our ansatz
\eq{3.4} reduces torsion just to its tensorial piece, it is not surprising
that for $\l=0$, the coefficient $\m^2$ in \eq{3.7} reduces exactly to
$m^2$. For $\l\ne 0$, $\m^2$ is associated to the spin-$2$ excitations
around the (A)dS background, and the condition for the absence of tachions
requires $\m^2\ge 0$.

In what follows, we will solve two dynamical equations \eq{3.7} for the
unknown functions $\cK$ and $G$, assuming $\m^2\ge 0$; then, we will use
\eq{3.3b} to find $S$. The torsion function $K$ and the metric function
$S$, obtained in this way, completely define the solution.

\section{Massive torsion waves}
\setcounter{equation}{0}

In this section, we classify the solutions of the field equations \eq{3.7}
for $\m^2>0$, according to the values of $\l$.

\subsection{\bm{\l=0}}

The simplest form of equations \eq{3.7} is obtained in the limit $\l\to 0$:
\bea
&&a_0G'-a_1K'=0\, ,\qquad \L=0\, ,                              \nn\\
&&K''+m^2K=0\, ,\qquad m^2=\frac{a_0(a_1-a_0)}{b_4 a_1}\,,      \lab{4.1}
\eea
with $G=S'+K$ and $S=H/2$. The solution has a simple form:
\bea
&&K=A(u)\cos my+B(u)\sin my\, ,                                 \nn\\
&&\frac{1}{2}H=
    \frac{a_1-a_0}{a_0 m}(A\sin my-B\cos my)+h_1(u)+h_2(u)y\, . \lab{4.2}
\eea

In Riemannian gravity, one can remove the term $h_1+h_2y$ in $H$ by a coordinate transformation. This transformation does not change the form of the metric \eq{2.1},
which is the only dynamical variable of the theory in vacuum. In the RC theory, such
a coordinate transformation is not particularly useful as it affects the form of the connection. Note, however, that the term $h_1+h_2y$ has no influence upon the RC
curvature, which depends only on $H''$. Thus, without loss of generality, we can
choose $h_1=h_2=0$.

The vector field $k=\pd_v$ is the Killing vector for both the metric and
the torsion; moreover, it is a null and covariantly constant vector field.
This allows us to consider the solution \eq{4.2} as a generalized pp-wave.

\subsection{\bm{\l>0}}

For positive $\l$, we use the notation
$$
\l=\frac{1}{\ell^2}\,,\quad x=\frac{y}{2\ell}\,,\quad \k=2\mu\ell\, ,
$$
so that $ \int dy=2\ell\int dx$. Now, having in mind the form of the
solution \eq{4.2} for $\l=0$, we use a similar ansatz for the torsion
function $\cK\equiv qK$:
\bsubeq\lab{4.3}
\be
\cK=A\cos\a+B\sin\a\,,\qquad \a=\a(y)\, ,
\ee
where $A=A(u),B=B(u)$. Substituting this into $(2ND)$ of \eq{3.7} produces
two conditions on $\a$:
$$
p^2(\a')^2-\mu^2=0\, ,\qquad p^2\a''-\frac{1}{2}\l yp\a'=0\, .
$$
The first condition yields
\be
\a'=\frac{\m}{p}=\frac{\m}{1+x^2}\quad\Ra\quad
\a=2\ell\int\frac{\m}{1+x^2}dx=\k\arctan x\, ,
\ee
\esubeq
whereas the second one is automatically satisfied. In the limit $\l\to
0$, we have $\a\to\k x=my$, and \eq{4.3} reduces to \eq{4.2}.

In the next step, we use \eq{4.3} and $(1ST)$ to calculate $G$:
\be
G=2\ell C_0\int\frac{q}{p}\cK'dx
   =D\frac{1}{p}\left[\left(q A-\frac{4x}{\k}B\right)\cos\a
    +\left(qB+\frac{4x}{\k}A\right)\sin\a\right]\, ,            \nn
\ee
where $D=C_0\k^2/(\k^2-4)$. Finally, integrating the relation
$S'=(p/q^2)G-K$ yields the metric function $H$. Using the definition
\be
\cH:=\frac{\sqrt{p}}{2q}H\equiv S+\frac{\l }{2}v^2,
\ee
we find:
\bea
\cH&=&\cH_1+\cH_2\,,                                            \nn\\[2pt]
\cH_1&:=&2\ell\int\frac{p}{q^2}Gdx
    =2\ell D\cdot\frac{p}{\k q}\left(A\sin\a-B\cos\a\right)
     \, ,                                     \nn\\[2pt]
\cH_2&:=&-2\ell\int Kdx=\frac{2\ell}{\k^2-4}\times              \nn\\[2pt]
&& \left[(B-iA)(2+\k)e^{i(2-\k)\arctan x}\,{}_2F_1
 \left(1,\frac{2-\k}{4};\frac{6-\k}{4};-e^{4i\arctan x}\right)\right. \nn\\
&&-\left.(B+iA)(2-\k)e^{i(2+\k)\arctan x}\,
                  {}_2F_1\left(1,\frac{2+\k}{4};\frac{6+\k}{4};
                              -e^{4i\arctan x}\right)\right]\,, \lab{4.4}
\eea
where ${}_2F_1(a,b;c;z)$ is the hypergeometric function \cite{x24}. Here, again, the integration term $h_1(u)+h_2(u)y/q$ appearing in $\cH$ is removed,
as it has no influence upon the RC curvature.

In order to illustrate the form of the torsion wave, we display here the
plots of the torsion function $(q^2/p)K(u,y)$ and the curvature function
$pG'(u,y)/2$, for a specific choice of the parameters $\ell,\k$, and for
fixed amplitudes $A(u)$ and $B(u)$.

\begin{figure}[htb]
\centering
\includegraphics[height=3.5cm]{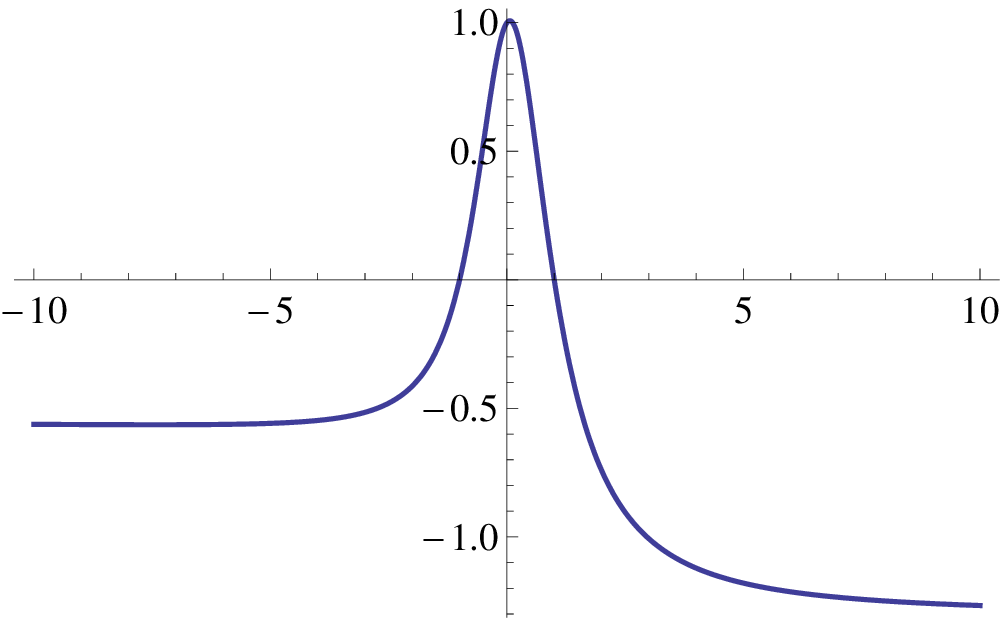}
   \qquad \includegraphics[height=3.5cm]{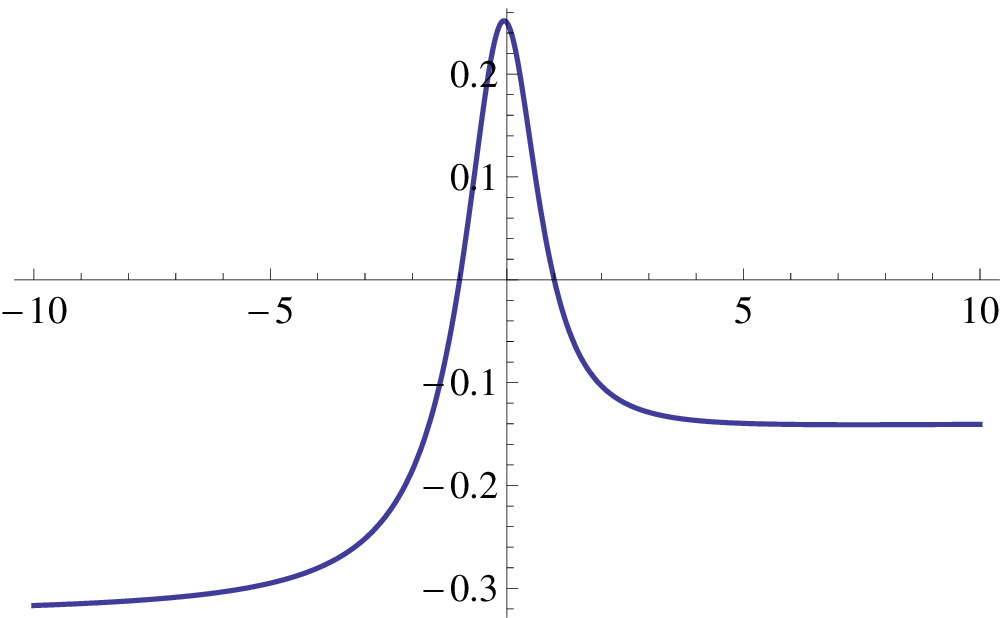} \\
\caption{The form of the torsion function $(q^2/p)K$ (left) and the
curvature function $pG'/2$ (right) for $\m^2>0$, in the region
$x\in[-10,10]$, and for $A(u)=B(u)=1$, $\ell=1$, $\k=1/4$.}\label{fig1}
\end{figure}

\subsection{\bm{\l<0}}

In this case, we use the notation
$$
\l=-\frac{1}{\ell^2}\, ,\quad x=\frac{y}{2\ell}\, ,\qquad \k=2\ell\m\,,
$$
and find that the torsion function $\cK$ is given by
\be
\cK=A\cos\a+B\sin\a\,,\qquad
\a=\k\,\frac{1}{2}\ln\left|\frac{1+x}{1-x}\right|=\k\atgh x\, .
\ee
Here, $\a(x)$ is singular at $x=1$, but for $\l\to 0$, it has the expected
limit: $\a\to\k x=my$. Then, following the same steps as in the previous
subsection, we can first calculate $G$,
\be
G=\frac{E}{p}\left[\left(Bq-\frac{4x}{\k}A\right)\sin\a
                  +\left(Aq+\frac{4x}{\k}B\right)\cos\a\right]\,,\nn
\ee
where $E=C_0\k^2/(\k^2+4)$, and then find the metric function $\cH$:
\bea
\cH&=&\cH_1+\cH_2\, ,                                           \nn\\[2pt]
\cH_1&:=&2\ell \int\frac{p}{q^2}Gdx
     =2\ell \frac{E}{\k}\frac{p}{q}\bigl[A\sin\a-B\cos\a\bigr]\, ,                                    \nn\\[2pt]
\cH_2&:=&-2\ell\int Kdx=-\frac{2\ell i}{\k^2+4}\times           \nn\\[2pt]
 &&\left[(B-iA)(2+i\k)e^{(2-i\k)\atgh x}\,
      {}_2F_1\left(1,\frac{2-i\k}{4};\frac{6-i\k}{4};
                                    -e^{4\atgh x}\right)\right. \nn\\
 &&-\left.(B+iA)(2-i\k)e^{(2+i\k)\atgh x}\,
      {}_2F_1\left(1,\frac{2+i\k}{4};\frac{6+i\k}{4};
                                    -e^{4\atgh x}\right)\right]\, .
\eea
As before, all the integration terms in $\cH$ are removed.

This solution can be obtained from the one for $\l>0$ by the analytic
continuation in $\ell$:
$$
\ell\to i\ell\quad\Ra\quad \k\to i\k,\quad x\to\frac{1}{i}x\,,\quad
\arctan x\to \frac{1}{i}\atgh x\, .
$$
For the asymptotic behavior of both massive and massless torsion waves,
see section \ref{Dis} and Appendix B.

\section{Massless torsion waves}
\setcounter{equation}{0}

For $\mu^2=0$, we have $a_1-a_0-b_6\l=0$ and the field equations \eq{3.7}
are simplified:
\be
pG'=C_0q\cK'\, ,\qquad p(p\,\cK')'=0\, .                        \lab{5.1}
\ee

\subsection{\bm{\l=0}}

For vanishing $\l$, the field equations with $C_0=1$ take the form:
\be
G'-K'\equiv\frac{1}{2}H''=0\, ,\qquad K''=0\, ,
\ee
so that
\be
 H=h_1(u)+h_2(u)y\,,\qquad K=k_1(u)+k_2(u)y\, .
\ee
This is a rather strange solution: since the metric function $H$ is
trivial, the metric takes the Minkowski form and consequently, it is
dynamically decoupled from the torsion.

\subsection{\bm{\l>0}}

\begin{figure}[htb]
\centering
\includegraphics[height=3.5cm]{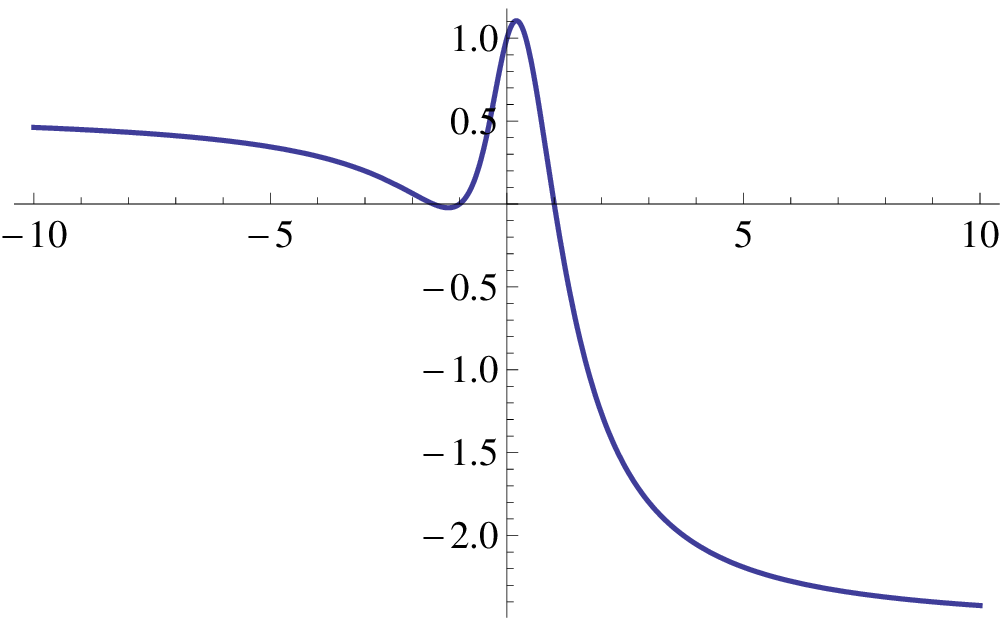}
   \qquad \includegraphics[height=3.5cm]{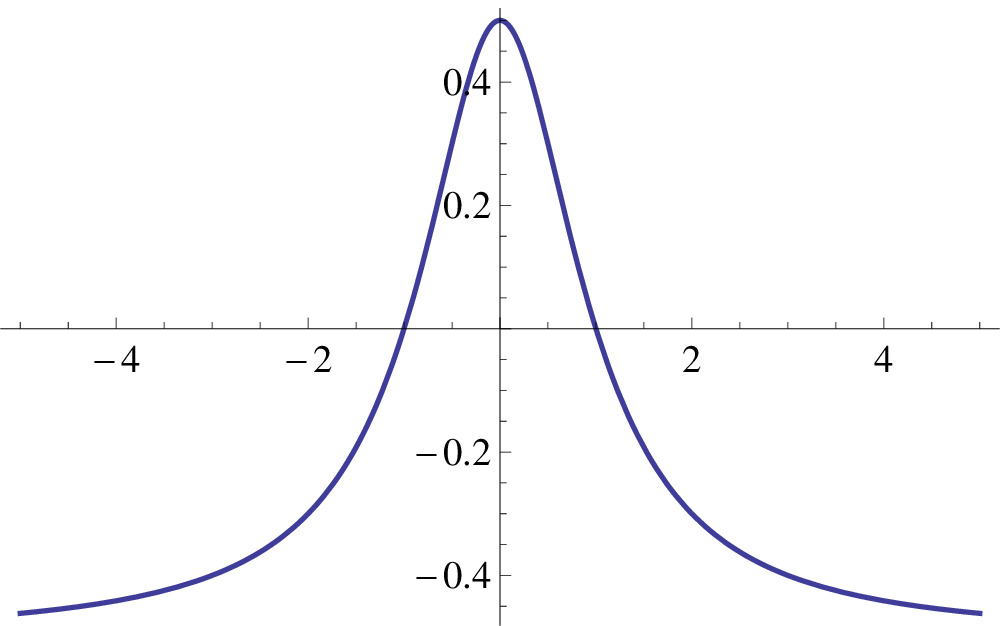} \\[9pt]
\caption{The form of the torsion function $(q^2/p)K$ (left) and the
curvature function $pG'/2$ (right) for $\m^2=0$, in the region
$x\in[-10,10]$, and for  $A(u)=B(u)=1$ and $\ell=1$.} \label{fig2}
\end{figure}

For the positive cosmological constant, with $\l:=1/\ell^2$ and
$x=y/2\ell$, the solution reads:
\bea
\cK&=&A(u)\arctan x + B(u)\,,                                   \nn\\
G&=&A(u)\frac{C_0x}{p}\,,                                       \nn\\
\cH(u,y)&=&\ell A(u)\left(\frac{C_0}q
  -\arctan x\cdot\ln\frac{1-ie^{2i\arctan x}}{1+ie^{2i\arctan x}}\right)\nn\\
  &&+\frac{i\ell}{2} A(u)\Biggl[\Li_2\Bigl(ie^{2i\arctan x}\Bigr)
                               -\Li_2\Bigl(-ie^{2i\arctan x}\Bigr)\Biggr]\nn\\
  &&-2\ell B(u)\atgh x,\qquad
\eea
where Li${}_2(z)$ is the dilogarithm function \cite{x24}. The solution is
illustrated in Fig. \ref{fig2}.

\subsection{\bm{\l<0}}

Finally, for $\l:=-1/\ell^2$, one finds:
\bea
\cK&=&A(u)\atgh x + B(u)\,,                                     \nn\\
  G&=&A(u)\frac{C_0x}{p}\,,                                     \nn\\
\cH(u,y)&=&\ell A(u)\left(
 -\frac{C_0}q-\frac i2\atgh x\cdot
  \ln\frac{1-ie^{-2\atgh x}}{1+ie^{-2\atgh x}}\right)           \nn\\
&&+\frac{i\ell}{2}A(u)\Biggl[ \Li_2\Bigl(ie^{-2\atgh x}\Bigr)
    -\Li_2\Bigl(-ie^{-2\atgh x}\Bigr)\Biggr]                    \nn\\
&&+2\ell B(u)\arctan x\, .
\eea

\section{Discussion and conclusions}\label{Dis}
\setcounter{equation}{0}

In this paper, we derived a new class of exact solutions of 3D gravity
with propagating torsion in empty spacetime, the generalized plane-fronted
waves, or the torsion waves.

The wave ansatz for the metric \eq{2.4} and the RC connection \eq{3.3}
represents a natural generalization of the Riemannian plane-fronted waves
with cosmological constant. However, a covariant characterization of the
wave nature of an exact solution is a rather complex issue
\cite{x13,x14,x15}, which has not been fully clarified for non-Riemannian
theories of gravity; for an attempt in this direction, see \cite{x25}.

The existence of the null covector $k_i=(1,1,0)$, appearing already in the
RC connection \eq{3.3}, is an essential element of the geometric structure
of a gravitational wave. It can be represented as the 1-form $k_i
b^i=\sqrt{2}du$, associated to the wave fronts $u=$ const. The related
vector field $k^i\pd_i=\sqrt{2}\pd_v$ is orthogonal to the $y$ direction;
moreover, for $\l=0$, $k^i$ is covariantly constant (pp-wave).

Based on an analogy with the electromagnetism, Lichnerowicz proposed a
covariant criterion for the existence of gravitational waves in general
relativity, see \cite{x14}. After separating the radiation piece of the RC
curvature \eq{3.5}, $S^{ij}:=R^{ij}+2\l b^ib^j$,
one can verify that it satisfies Lichnerowicz's requirements:
\be
k^iS_{ijmn}=0\, , \qquad \ve^{ijk}k_iS_{jkmn}=0\, .             \lab{6.1}
\ee
Clearly, the above criterion is not sufficient for a RC geometry, where we
have one more field strength, the torsion. However, in analogy with
electromagnetism, the radiation conditions for torsion are expected to
have the form
\be
k^m T_{imk}=0\, ,\qquad \ve^{mnk}k_mT_{ink}=0\, .               \lab{6.2}
\ee
A direct verification based on \eq{3.4} shows that these conditions are
also satisfied. The radiation properties \eq{6.1} and \eq{6.2} strongly
support the interpretation of our generalized plane-fronted wave as a
genuine PGT extension of the related Riemannian structure.

One should also note that our RC curvature has the same irreducible
components as the corresponding Riemannian curvature, and moreover, it has
all the usual index symmetries of the Riemannian curvature; in particular,
$R_{ijmn}=R_{mnij}$. The same properties were found by Pa\v si\'c and
Vassiliev \cite{x26} in their pp-wave with torsion, constructed in the
model with metric-compatible connection and curvature squared Lagrangian.
The torsion of their solution is pure tensor, as in our case.

In electrodynamics and in general relativity, exact wave solutions are
associated to massless modes of the related fields, so that the appearance
of massive torsion waves may seem a bit strange. However, the existence of
massive torsion modes is not in conflict with the gauge structure of PGT,
it is a generic feature associated to the presence of $T^2$ terms in the
Lagrangian. Massive waves appear also in some Riemannian extensions of GR,
such as Topologically massive gravity or New massive gravity
\cite{x19,x20,x21}.

Asymptotic properties of the torsion waves are defined by the large $y$
limits of the torsion \eq{3.4} and the RC curvature \eq{3.5}. As follows
from the results of Appendix B, the generic asymptotic form of the torsion
waves does not coincide with the (A)dS geometry.

Our study of exact torsion waves in 3D can be considered as a complement
to the related results in four dimensions \cite{x25,x26,x27}. In particular,
we wish to place emphasis on the results of Sippel and Goenner \cite{x25},
who made a significant progress in clarifying the structure of pp-waves
with torsion: (i) they generalized the Ehlers--Kundt classification of
pp-waves \cite{x13} by relaxing the assumption that the GR field equations
hold, and (ii) they introduced a classification of the allowed form of
torsion in pp-waves. Further advances in this direction would help us to
better understand the role of torsion in exact wave solutions.

\section*{Acknowledgements}

One of us (M.B.) would like to thank Yuri Obukhov for useful comments.
This work was supported by the Serbian Science Foundation under Grant
No. 171031.

\appendix
\section{Irreducible decomposition}
\setcounter{equation}{0}

For the sake of completeness, we present here the form of the irreducible
components of $T^i$ and $R^{ij}$, see also \cite{x8}, with the wedge
product sign explicitly displayed.

Torsion has three irreducible components, the vector, axial and tensor
component:
\bea
&&{}^{(2)}T_i:=\frac{1}{2}b_i\wedge(h_m\hook T^m)
              =\frac{1}{2}\eta_{ij}V_k b^j\wedge b^k \, ,       \nn\\
&&{}^{(3)}T_i:=
  \frac{1}{3}\hd\bigl[b_i\wedge\hd(T^m\wedge b_m)\bigr]
             =\frac{1}{2}A\ve_{ijk}b^j\wedge b^k\,,             \nn\\
&&{}^{(1)}T_i:=T_i-{}^{(2)}T_i-{}^{(3)}T_i\, ,
\eea
where $V_k:=T^m{}_{mk}$ and $A:=\ve_{ijk}T^{ijk}/6$.

The curvature also has three irreducible pieces. Making use of the
definitions
\bea
&&A_i:=\frac{1}{2}h_i\hook\big(b^k\wedge\hR_k\big)
      =\hR_{[ik]}b^k\, ,                                        \nn\\
&&\sS_i:=\hR_i-A_i-\frac{1}{3}Rb_i=\hR_{(ik)}b^k-\frac{1}{3}Rb_i\, ,                                \nn
\eea
where $\hR_i:=\ric_i$, the irreducible pieces of $R_{ij}$ read:
\bea
&&{}^{(4)}R_{ij}:=b_i\sS_j-b_j\sS_i\, ,                         \nn\\
&&{}^{(5)}R_{ij}:=b_iA_j-b_j A_i\, ,                            \nn\\
&&{}^{(6)}R_{ij}:=\frac{1}{6}Rb_i\wedge b_j\, .
\eea
Note that in 3D, the Weyl curvature vanishes.

\section{Asymptotic geometry}
\setcounter{equation}{0}

In this appendix, we calculate the large $y$ limits of the expressions
$(q^2/p)K$ and $pG'/2$; these limits define the respective asymptotic
values of the torsion and the radiation piece of the curvature,
characterizing the gravitational wave. The formulas for $\l=0$ are
omitted, as the related asymptotic behavior can be read off directly from
the main text.

\subsection*{Case \bm{\m^2>0}}

\noindent{$\l>0$:}
\bea
&&\lim_{y\rightarrow\pm\infty}\frac{q^2}{p}K
  =-\left(A\cos\frac{\k\pi}{2}\pm B\sin\frac{\k\pi}{2}\right)\,,\nn\\
&&\lim_{y\rightarrow\pm\infty}\frac{1}{2}pG'
  =\frac{1}{2}C_0\mu \left(\pm A\sin\frac{\k\pi}{2}-B\cos\frac{\k\pi}{2}\right)\,.
\eea

\noindent{$\l<0$:}
\bea
&&\lim_{y\rightarrow \infty}\frac{q^2}{p}K=-A\, ,               \nn\\
&&\lim_{y\rightarrow \infty}\frac12pG'=-\frac 12 C_0\mu B\,.
\eea

\subsection*{Case \bm{\m^2=0}}

\noindent{$\l>0$:}
\bea
&&\lim_{y\rightarrow \pm\infty}\frac{q^2}{p}K
                        =\mp A\frac{\pi}{2}- B\, ,              \nn\\
&&\lim_{y\rightarrow \infty}\frac{1}{2}pG'=-\frac{AC_0}{4\ell}\, .
\eea

\noindent{$\l<0$:}
\bea
&&\lim_{y\rightarrow \infty}\frac{q^2}{p}K=-B\,,\nn\\
&&\lim_{y\rightarrow \infty}\frac12pG'=-\frac{AC_0}{4\ell}\,.
\eea



\begin{thebibliography}{99}

\bibitem{x1} For a review and an extensive list of references, see: S.
    Carlip, \emph{Quantum Gravity in 2+1 Dimensions} (Cambridge University
    Press, Cambridge, 1998); Conformal field theory, (2+1)-dimensional
    gravity, and the BTZ black hole, Class. Quantum Grav. {\bf 22},
    R85--R124 (2005).

\bibitem{x2} E. W. Mielke and P. Baekler, Topological gauge model of
    gravity with torsion, Phys. Lett. A {\bf 156}, 399--403 (1991).

\bibitem{x3} T. W. B. Kibble, Lorentz invariance and the gravitational
    field, J. Math. Phys. {\bf 2}, 212--221 (1961); D. W. Sciama, On the
    analogy between charge and spin in general relativity, in:
    \emph{Recent Developments in General Relativity, Festschrift for
    Infeld} (Pergamon Press, Oxford; PWN, Warsaw, 1962) pp. 415--439.

\bibitem{x4} For a textbook exposition of PGT, see: M. Blagojevi\'c,
    \emph{Gravitation and Gauge Symmetries} (Institute of Physics,
    Bristol, 2002); T. Ort\'{\i}n, \emph{Gravity and Strings} (Cambridge
    University Press, Cambridge, 2004).

\bibitem{x5} An up-to-date status of PGT, including its 3D version, can be
    found in: M. Blagojevi\'c and F. W. Hehl (eds.), \emph{Gauge Theories
    of Gravitation, A Reader with Commentaries} (Imperial College Press,
    London, 2013).

\bibitem{x6} Yu. N. Obukhov, Poincar\'e gauge gravity: Selected topics,
    Int. J. Geom. Meth. Mod. Phys. {\bf 3}, 95--138 (2006).

\bibitem{x7} J. A. Helay\"el-Neto, C. A. Hernaski, B. Pereira-Dias, A. A.
    Vargas-Paredes, and V. J. Vasquez-Otoya, Chern-Simons gravity with
    (curvature)$^2$- and (torsion)$^2$-terms and a basis of
    degree-of-freedom projection operators, Phys. Rev. D {\bf 82}, 064014
    (2010).

\bibitem{x8} M. Blagojevi\'c and B. Cvetkovi\'c, 3D gravity with
    propagating torsion: the AdS sector, Phys. Rev. D {\bf 85}, 104003
    (2012) [10 pages].

\bibitem{x9} M. Blagojevi\'c, B. Cvetkovi\'c, O. Mi\v skovi\'c and R.
    Olea, Holography in 3D AdS gravity with torsion, JHEP {\bf 1305}
    (2013) 103 [29 pages];

\bibitem{x10} M. Blagojevi\'c, B. Cvetkovi\'c, 3D gravity with propagating
    torsion: Hamiltonian structure of the scalar sector, Phys. Rev. D {\bf
    85}, 104032  (2013) [15 pages].

\bibitem{x11} H. Stephani, {\it Relativity, An Introduction to Special and
    General Relativity} (Cambridge University Press, Cambridge, UK, 2004),
    chapter 29.

\bibitem{x12} A. Peres, Some gravitational waves, Phys. Rev. Lett. {\bf
    3}, 571--572 (1959); reprinted in arXiv:hep-th/0205040.

\bibitem{x13} J. Ehlers and W. Kundt, Exact solutions of the gravitational
    field equations, in: \emph{Gravitation: an introduction to current
    research} (Willey, N.Y., 1962), ed. L. Witten, pp. 49--101.

\bibitem{x14} V. Zakharov, \emph{Gravitational waves in Einstein's theory}
    (Halsted Press, New York, 1973). 

\bibitem{x15} H. Stephani, D. Kramer, M. MacCallum, C. Hoenselaers, and E.
    Herlt, \emph{Exact solutions of Einstein's field equations}, 2nd ed.
    (Cambridge Univ. Press, Cambridge, 2003).

\bibitem{x16} Y. Obukhov, New solutions in 3D gravity, Phys. Rev. D
    {\bf 68} 124015 (2003).

\bibitem{x17} S. Deser, R. Jackiw and S.-Y. Pi, Cotton blend gravity pp
    waves, Acta Phys. Polon. B {\bf 36}, 27--34 (2005).

\bibitem{x18} E. Ay\'on-Beato and M. Hassa\"{\i}ne, pp-waves of conformal
    gravity with self-interacting source,  Annals Phys. {\bf 317},
    175--181 (2005);

\bibitem{x19}  E. Ay\'on-Beato and M. Hassa\"{\i}ne, Scalar fields
    nonminimally coupled to pp-waves, Phys. Rev. D {\bf 71}, 084004
    (2005).

\bibitem{x20}  E. Ay\'on-Beato,  G. Giribet, and M. Hassa\"{\i}ne, Bending
    AdS waves with New Massive Gravity, JHEP {\bf 0905} (2009) 029.

\bibitem{x21} T. Moon and Y. S. Myung,  Polarization modes of
    gravitational waves in three-dimensional massive gravities, Phys. Rev.
    D {\bf 85}, 027501 (2012).

\bibitem{x22} I. Osv\'ath, I. Robinson and K. R\'ozga, Plane-fronted
    gravitational and electromagnetic waves in spaces with cosmological
    constant, J. Math. Phys. {\bf 26}, 1755--1761 (1985).

\bibitem{x23} Y. Obukhov, Generalized plane-fronted gravitational waves in
    any dimension, Phys. Rev. D {\bf 69}, 024013 (2004).

\bibitem{x24} The field equations \eq{3.6} for the torsion waves are
    checked using the \emph{Excalc package} of the computer algebra system
    \emph{Reduce}; after being transformed to the form \eq{3.7}, they are
    solved with the help of \emph{Wolfram Mathematica};  see also:\\
    \emph{Pocketbook of Mathematical Functions}, abridged edition of
    \emph{Handbook of Mathematical Functions}, M. Abramowitz and I. Stegun
    (eds.), material selected by M. Danos and F. Rafelski (Verlag Harri
    Deutsch, Frankfurt am Mein, FRG, 1984); chapters 15 (hypergeometric
    functions) and 27.7 (dilogarithm).

\bibitem{x25} R. Sippel and H. Goenner, Symmetry classes of pp-waves,
    Gen. Rel. Grav. {\bf 18}, 1229--1243 (1986).

\bibitem{x26} V. Pa\v si\'c and D. Vassiliev, PP-waves with torsion and
    metric-affine gravity,  Class. Quant. Grav. {\bf 22} 3961--3976
    (2005).

\bibitem{x27} W. Adamowicz, Plane waves in gauge theories of gravitation,
    Gen. Rel. Grav. {\bf 12}, 677--691 (1980); P. Singh and J. B.
    Griffiths, A new class of exact solutions of the vacuum quadratic
    Poincar\'e gauge field theory, Gen. Rel. Grav. {\bf 22}, 947--956
    (1990); V. V. Zhytnikov, Wavelike exact solutions of $R+R^2+Q^2$
    gravity, J. Math. Phys. {\bf 35}, 6001--6017 (1994); M.-K. Chen, D.-C.
    Chern, R.-R. Hsu, and W. B. Yeung, Plane-fronted torsion waves in a
    gravitational gauge theory with a quadratic Lagrangian, Phys. Rev. D
    {\bf 28}, 2094--2095 (1983); O. V. Babourova, B. N. Frolov and E. A.
    Klimova, Plane torsion waves in quadratic gravitational theories,
    Class. Quant. Grav. {\bf 16}, 1149--1162 (1999). The wave solutions
    found in the last two references are rather peculiar: in both cases,
    the solutions for the metric and the torsion are dynamically
    decoupled.

\end{thebibliography}
\end{document}